\documentclass[]{emulateapj}
\usepackage{natbib}

\def\ie{i.e.}
\def\eg{{e.g.}}
\def\ltsima{$\; \buildrel < \over \sim \;$}
\def\simlt{\lower.5ex\hbox{\ltsima}}
\def\gtsima{$\; \buildrel > \over \sim \;$}
\def\simgt{\lower.5ex\hbox{\gtsima}}

\def\eq{equation}
\def\pop3{Population~III}

\def\fig{Figure}

\def\hide#1{}

\received{.......}
\accepted{.......}
\slugcomment{to be submitted to ApJ}

\begin{document}
\markright{DRAFT: \today\hfill}

%\def\placefig#1{#1}
%\title{The peculiar microlensing magnification curve of primordial black holes as MACHOs}
%\title{Ultracompact Minihalos as MACHOs: an observational test}
\title{A new probe of dark matter and high-energy Universe using microlensing}
%\title{A new type of dark matter Machos}
\author{Massimo Ricotti\altaffilmark{1} \& Andrew Gould\altaffilmark{2}}
\altaffiltext{1}{Department of Astronomy, University of Maryland,\\
College Park, MD 20742. E--mail: ricotti@astro.umd.edu}
\altaffiltext{2}{Department of Astronomy Ohio State University,\\
Columbus, OH 43210.  E--Mail: gould@stronomy.ohio-state.edu}
%  .................................................................
 
\begin{abstract}
  We propose the existence of ultracompact minihalos as a new type of
  massive compact halo object (MACHO) and suggest an observational
  test to discover them. These new MACHOs are a powerful probe into the
  nature of dark matter and physics in the high energy
  Universe.  

  Non-Gaussian energy-density fluctuations produced at phase
  transitions (\eg, QCD) or by features in the inflation potential can
  trigger primordial black hole (PBH) formation if their amplitudes
  are $\delta \simgt 30\%$. We show that a PBH accumulates over time a
  sufficiently massive and compact minihalo to be able to modify or
  dominate its microlensing magnification light curve. Perturbations
  of amplitude $0.03\% \simlt \delta \simlt 30\%$ are too small to
  form PBHs, but can nonetheless seed the growth of ultracompact
  minihalos. Thus, the likelihood of ultracompact minihalos as MACHOs
  is greater than that of PBHs. In addition, depending on their mass,
  they may be sites of formation of the first \pop3 stars.

  Ultracompact minihalos and PBHs produce a microlensing light curve
  that can be distinguished from that of a ``point-like'' object if
  high-quality photometric data are taken for a sufficiently long time
  after the peak of the magnification event. This enables them to be
  detected below the stellar-lensing ``background'' toward both the
  Magellanic Clouds and the Galactic bulge.

\end{abstract}

\section{Introduction}\label{sec:int}

The nature of dark matter is one of the fundamental unsolved questions
in Cosmology. CMB anisotropy data indicate that about $20\%$ of the
Universe is composed of non-baryonic dark matter and $4\%$ is in
baryons. The most popular dark matter candidates are weakly
interactive massive particles (WIMPs) and a great effort is under way to
detect them. Ongoing experiments using ground based dark matter
detectors are significantly restricting the parameter space for some
favored supersymmetric models \citep[\eg,][]{WIMP:08, WIMP:09}.

A less popular -- but physically well-motivated -- dark matter
candidate is primordial black holes (PBHs). These are black holes
predicted to form during the radiation dominated era if the Universe
had regions with energy-density fluctuations of amplitude $\simgt
30\%$. PBHs can form with a range of masses that span many decades,
from the Planck mass ($10^{-5}$ g) to thousands of solar masses. 

A fully relativistic approach is necessary to model the formation of
PBHs. However, the Newtonian approximation elucidates the basic
concept of why PBHs can form before large scale structures appeared in
the Universe.  During the radiation era the cosmic Jeans mass
approaches the Horizon mass, which is also of the same order of the
mass of a black hole with density equal to the mean cosmic value. This
means that before matter-radiation equality relatively small
perturbations on Horizon scales may become self gravitating and they
begin the collapse with an initial density that is already very close
to the black hole regime. Thus, the mass of PBHs depends on the time
of their formation because it is of the order of the mass of the
particle Horizon at that time.

However, fluctuations of amplitude of about $30\%$ - required for PBH
formation - are very large when compared to the r.m.s amplitude of
perturbations from inflation (about $0.001\%$) that lead to the
formation of normal galaxies.  Thus, such large perturbations would be
exceedingly rare, unless they can be produced by physical processes
taking place during phase transitions (\eg, topological defects,
bubble nucleation, softening of the cosmic equation of state, etc) or
by ad hoc features in the inflation potential. Due to theoretical
uncertainties in the physics of the high-energy Universe we do not
know the typical amplitude of perturbations created at phase
transitions. Hence, theoretical modeling cannot tell us whether PBHs
exist, whether they are the bulk of dark matter or whether they are
only an academic curiosity. We must rely on observations to probe them
as dark matter candidates.

Even if PBHs are only a fraction (\eg, $<10\%$) of the dark matter,
during matter domination, they can grow by up to two orders of
magnitude in mass through the acquisition of large dark matter halos
\citep[][hereafter MOR07]{MackOR:07}. The dark halo is instrumental in
increasing the ability of the PBHs to accrete gas, in boosting their
X-ray emission, and thus in modifying the cosmic ionization history
\citep[][hereafter ROM08]{RicottiOM:08}.  In this paper we consider
whether the dark halo is sufficiently massive and compact to produce
observable modifications of microlensing light curves of PBHs or even
dominate the inferred mass of the lens. Most importantly, we propose a
novel type of non-baryonic MACHO. The idea is a corollary of models of
PBH formation: if the amplitude of the fluctuations that may induce
collapse of PBHs is smaller than the required threshold of $\sim
30\%$, then even though the PBH is not formed, the perturbation will
nonetheless seed the growth of ultracompact minihalos observable using
microlensing experiments. Due to the lower overdensity threshold
required to form ultracompact minihalos, their existence is several
orders of magnitude more likely than that of PBHs.

Observational evidence for (or constraints against) PBHs is often
sought in conflicting claims for (MACHO: \citealt{Alcock:00}) or
against (EROS: \citealt{Tisserand:07}; OGLE: \citealt{Wyrzykowski:09})
detection of massive compact halo objects (MACHOs) in microlensing
experiments toward the Magellanic Clouds (MCs). For example,
\citealt{Alcock:00} originally claimed that roughly 20\% of Galactic
dark matter was in the form of MACHOs of characteristic mass
$0.4\,M_\odot$, which made PBHs a good candidate (since stars of this
mass would easily be seen). However, ROM08 argued that such a
population of PBHs would emit X-rays and produce distortion of the CMB
spectrum that are incompatible with COBE-FIRAS data.

Here, however, we adopt a very different orientation.  First, we
reopen the possibility that PBHs can be identified with microlenses
detected toward the MCs by noting that the PBH contribution to the
microlens mass is small compared to that of accreted dark matter.
Hence the X-ray signature is likewise small.  Second, we argue
that these PBH-seeded compact objects can have most of their mass
outside the lensing Einstein radius, meaning that the objects
might be cosmologically very important even if naive interpretations
of the lensing results constrain their density to be $<10\%$ of
Galactic dark matter (which would be consistent with both 
experiments claiming upper limits).  Third, and most important,
we argue that because they are embedded in accreted dark matter,
PBHs generate a lensing signal that is qualitatively different
from that due to stars and other ``point-like'' objects.  Hence,
PBHs would be detectable (and distinguishable from stars) 
in microlensing experiments even if their lensing rate was
equal to or below that of stars.  Moreover, even if the amplitude
of early-universe perturbations is too small to create PBHs, it may
still be large enough to generate compact minihalos that would
give rise to lensing signatures.  Like the PBH-minihalo signatures,
these could also be robustly distinguished from garden-variety
microlensing due to stars.  Thus, we propose an observational
test for PBHs and other early-universe perturbations that is
far more sensitive and robust than any previously contemplated.

The plan of the paper is as follows. In \S~\ref{sec:halo} we discuss
the formation and derive the density profile of ultracompact
minihalos. In \S~\ref{sec:mag} we calculate the magnification light
curve of microlensing events due to PBHs embedded in their minihalos
and ultracompact minihalos without PBHs. In \S~\ref{sec:bulge} we
discuss the prospects of detecting PBHs and ultracompact minihalos
below the stellar-lensing ``background'' toward both the Magellanic
Clouds and the Galactic bulge. Finally, in \S~\ref{sec:conc} we
present the discussion and conclusions.

\section{Density Profile of  Ultracompact Minihalos}\label{sec:halo}

It is generally accepted that the mass of PBHs does not increase
significantly after their formation \citep{Zeldovich:67, CarrH:74}.
However, any locally overdense region in an expanding universe seeds
the formation of dark matter structures. PBHs are local overdensities
in the dark matter distribution (made either of WIMPs or smaller mass
PBHs), hence they seed the growth of spherical halos.

The theory of spherical gravitational collapse in an expanding
universe (\ie, assuming radial infall), also known as secondary infall
theory \citep{Bertschinger:85}, predicts that during the
matter-dominated era, the dark halo grows as $t^{2/3} \propto
(1+z)^{-1}$.  During the radiation-dominated era the halo growth is of
order unity, thus the halo mass grows with redshift as
\begin{equation}
M_{h}(z)=M_{pbh}\left(1+z \over 1+z_{eq}\right)^{-1},
\label{eq:mhalo}
\end{equation}
where $z_{eq} \approx 3500$ is the redshift of matter-radiation
equality (MOR07). After $z \sim 30$, the growth of the dark minihalo
depends on the environment. If the PBH evolves in isolation it can
continue to grow, otherwise it will either stop growing or it will
lose mass due to tidal interactions as it is incorporated into a
larger galactic halo.

\subsection{Revisiting Secondary Infall Seeded by PBHs}

In previous work on halo growth seeded by PBHs, MOR07 have assumed a
point-like overdensity (\ie, the PBH) accreting gas and dark matter
from a uniform density expanding universe with $\rho=\Omega_m
\rho_{crit}$ (the growth of the halo mass during radiation epoch is
negligible).  However, this assumption is correct only if the mass of
the perturbation that originates the PBH is equal to the PBH
mass. This is typically incorrect. We expect $M_{pbh} \approx
M_{Hor}$, only if the perturbation that creates the PBH has amplitude
much larger than the critical amplitude for collapse.

One dimensional fully-relativistic simulations of gravitation collapse
of perturbations on Horizon scales show that the critical overdensity
needed to trigger PBH collapse is $\approx w$, where $P=w \rho c^2$ is
the cosmic equation of state \citep{Carr:75, Green:04}. During the
radiation era $w=1/3 \sim 30\%$. The mass of PBHs is typically smaller
than the mass within the particle Horizon at the redshift of its
formation, $z_f$: $M_{pbh} =f_{Hor}M_{Hor}(z_f)$, where $f_{Hor}<1$
depends on the amplitude of the perturbation relative to the critical
value $w$. Perturbations with amplitude significantly larger than $w$
produce $f_{Hor} \sim 1$, while perturbation that are near the
critical value produce PBHs with masses much smaller than
$M_{Hor}$. In addition, the total mass of the perturbation can be
$\delta m > M_{Hor}(z_f)$ if it has wings of subcritical amplitude
extending outside of the Horizon at $z_f$. Partially for this reason
it has been suggested by several authors \citep[\eg,][]{Dokuchaev:04,
  Carr:05, Chisholm:06} that PBHs have high probability of forming
clusters or binaries.

According to the theory of secondary infall any mass excess $\delta
m>0$ within a sphere of radius $R$ seeds the growth of a
minihalo. Here $\delta m = M(R)-4\pi/3 \Omega_m\rho_{cr} R^3$ includes
the mass of the PBH and any overdensity surrounding it. We ran several
simulations (see Section~\ref{ssec:runs}) in which the PBH is located
in an overdense region confirming that the mass of the minihalo is
proportional to the mass of the overdense region $\delta m$:
 \begin{equation}
M_{h}(z)=\delta m \left(1+z \over 1+z_{eq}\right)^{-1}.
\label{eq:mhalo_new}
\end{equation} 
If the accretion has spherical symmetry (\ie, radial infall), the dark
halo develops a self-similar power-law density profile $\rho \propto
R^{-\nu}$ with $\nu \sim 2.25$ \citep{Bertschinger:85} truncated at a
halo radius
\begin{equation}
R_{h} = 0.019~{\rm pc}\left({M_{h}(z) \over
1~M_\odot}\right)^{1/3}\left({1+z \over 1000}\right)^{-1},
\label{eq:rhalo}
\end{equation}
where $R_h$ is about one third of the turn-around radius
\citep[][ROM08]{Ricotti:07}.  Hence, the cumulative mass of the halo
is $M_h(R)=M_h(z)(R/R_h(z))^{3/4}$.

\subsubsection{Simulations}\label{ssec:runs}

\begin{figure*}[tbh]
\plottwo{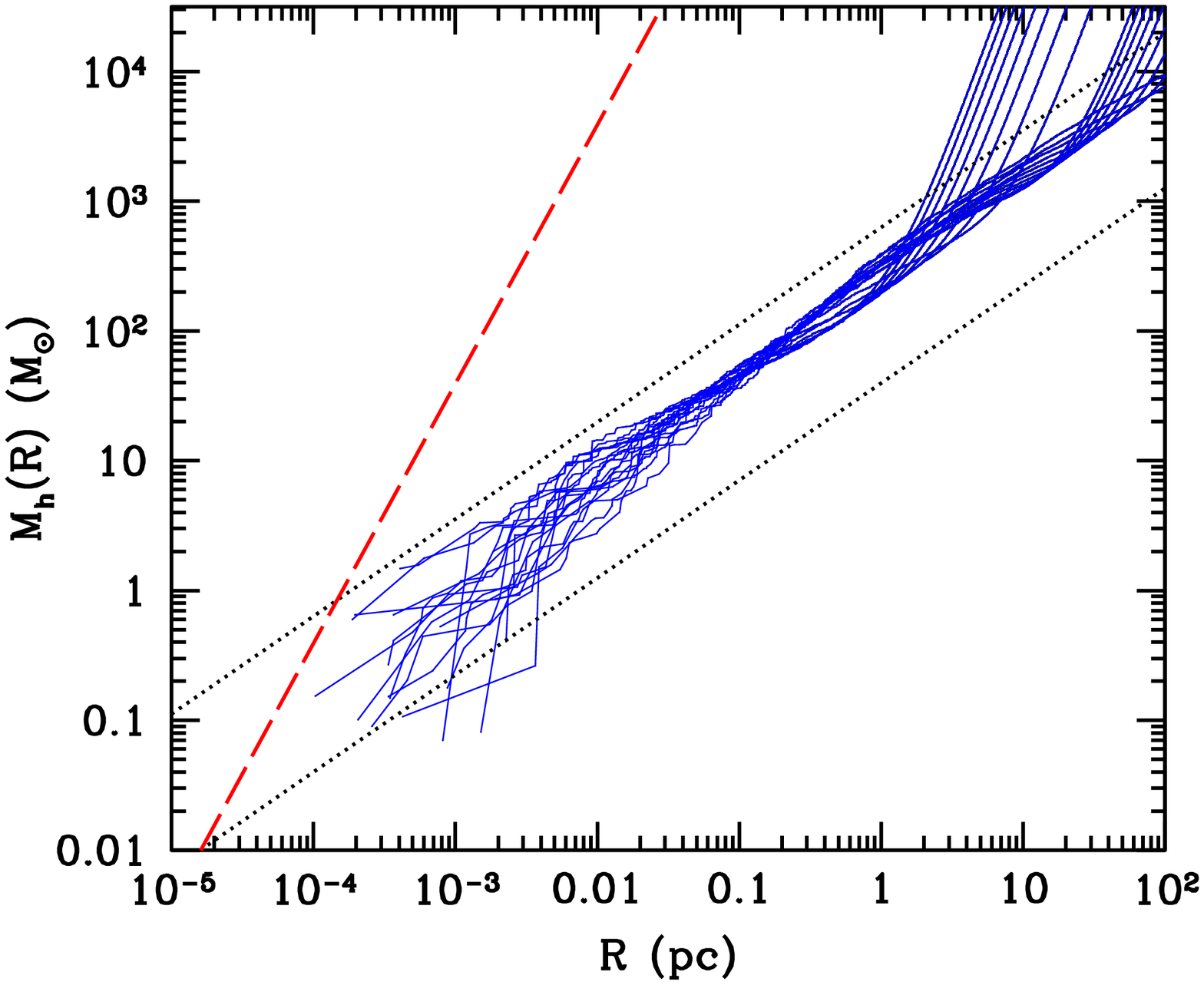}{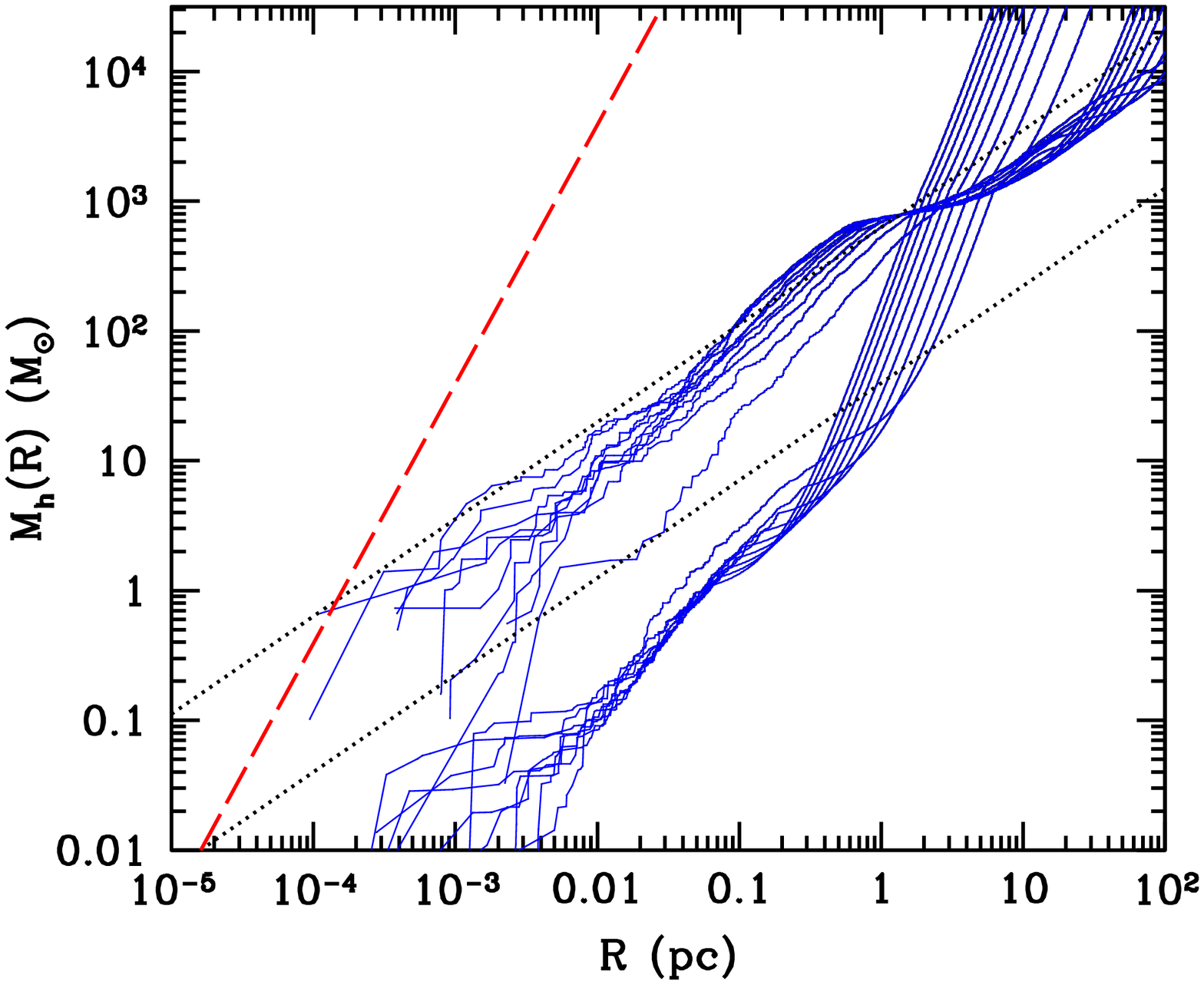}
\caption{\label{fig:compr}{\it (Left)} Time sequence of the growth of
  the mass profile of a spherical halo due to secondary infall. The
  halo growth is seeded by a density perturbation with overdensity
  $\delta(z_{eq}) =10$ and excess mass $\delta m =100$~M$_\odot$. The solid
  lines show the time evolution of the mass profile from $z=1000$ to
  $z=5$. The long dashed line delineate the lens equation. The two dotted
  lines show power law mass profiles with log slope $0.75$ producing
  gravitational lenses with $M_{lens}=0.01, 1$~M$_\odot$. {\it
    (Right)} Same as the left figure for a perturbation
  $\delta(z_{eq})=0.1$ and $\delta m=100$~M$_\odot$.}
\end{figure*}

We have run 1D simulations of secondary accretion that confirm the
results expected from the theory of secondary infall. The code we used
is a modified version of the code described in MOR07. The modification
allows for shell crossing and crossing of the singularity at $R=0$.

We ran simulations with different shape of the overdensity (top-hat,
triangular, etc), different overdensity, and different size. After a
transient phase the density profile always settles to a power law with
slope $2.25$. The mass profile of the halo depends on $\delta m$ as in
equation~(\ref{eq:mhalo_new}). However, if the initial amplitude of
the perturbation at equality is too small for the perturbation to grow
to $\delta \sim 1$ by a given redshift, the collapse has 2 phases:
during the first phase the halo is less dense and in the second phase,
after $\delta \sim 1$ the density increases to its final value. We
estimate the critical amplitude required for the formation of an
ultracompact minihalo as the one that leads to the second and final
phase of the collapse at $z\sim 1000$. This is because halos that
reach the final density profile at lower redshifts are likely to have
a shallower inner core due to the larger angular momentum of accreted
dark matter leading to some degree of non-radial infall.

\subsubsection{Lensing}

The microlensing magnification by the minihalo and PBH depends on the
mass profile projected on a plane. In Section~\ref{sec:mag} we find
that $M_{\perp}(R) = 1.350 M_h(R)$ for a density profile with
$\nu=2.25$. Hence, from equations~(\ref{eq:mhalo})-(\ref{eq:rhalo}):
\begin{equation}
M_{\perp}(R)=0.058~M_\odot \left({\delta m \over 1~M_\odot}\right)^{{3/4}}\left({R \over 8~{\rm AU}}\right)^{3/4}.
\label{eq:proj}
\end{equation}
The reason that we parameterize the radius in equation~(\ref{eq:proj})
in units of $8$~AU is that the Einstein radius of a lens with mass
$M_{lens}$ is about $R_{\rm E} \approx 8~{\rm AU} (M_{lens}/1~M_\odot)^{1/2}$. The
mass of the lens is $M_{lens}\approx M_{\perp}(R_E)+M_{pbh}$.  If
$M_{\perp}(R_E) \ll M_{pbh}$, then $M_{lens} \approx M_{pbh}$ and
\begin{equation}
{M_{\perp}(R_E) \over M_{pbh}}=5.8\% \left({\delta m \over 1~M_\odot}\right)^{3/4}\left({M_{pbh}\over 1~M_\odot}\right)^{-5/8}.
\end{equation}
In Section~\ref{sec:mag} we will show that if $M_{\perp}(R_E)\simgt
30\%M_{pbh}$ the difference of the light curve of the lensing event
from that of a point mass with $M_{lens}=M_{pbh}$ is measurable. If
$\delta m=M_{pbh}$ the effect of the minihalo in modifying the
magnification light curve of the PBH is probably not observable assuming
realistic measurement errors. However, we have already pointed out
that typically $\delta m \sim M_{Hor} > M_{pbh}$. If $\delta
m/M_{pbh}>9$ (or $f_{Hor}<0.11$ assuming $\delta m=M_{Hor}$) the
signature of the minihalo is observable in the lensing light curve.

In the limit of $M_{\perp}(R_E) \gg M_{pbh}$, then $M_{lens} \approx
M_{\perp}(R_E)$ and we find
\begin{equation}
M_{lens} = 1~M_\odot \left({\delta m \over 44.5~M_\odot}\right)^{6/5}.
\label{eq:halolens}
\end{equation}
In this case the mass of the lens is independent of the PBH mass. In
the next section we consider the case in which the PBH does not form:
$M_{pbh}=0$. This case is particularly interesting because the
formation of ultracompact minihalos is statistically more likely than
the formation of PBHs.

\subsection{Ultracompact Minihalos without PBHs}

Dark matter density fluctuations that seed the formation of large
scale structure and galaxies, have an amplitude $\delta \sim 10^{-5}$
when they enter the Horizon, roughly independently of their mass (for
scale invariant perturbations).  Here we focus on small mass
fluctuations that enter the Horizon during the radiation era. The
relativistic component of the (adiabatic) perturbations stops growing
rather quickly after it enters the Horizon because its mass becomes
smaller than the Jeans mass ($M_J(z) \sim M_{Hor}(z)$ for
$z>z_{eq}$). The dark matter component grows only logarithmically
because of the Meszaros effect. Overall, small mass linear
perturbations grow by 2-3 orders of magnitude from the time they enter
the Horizon to matter-radiation equality, depending on their mass. A
relatively rapid initial growth when the perturbation enters the
Horizon is followed by a slow logarithmic growth. For the mass range
of interest to us, the Horizon mass increases by 10 or more orders of
magnitude from the time the perturbation enters the Horizon to to the
redshift of equality ($M_{Hor}(z_{eq}) \sim 10^{15}$~M$_\odot$).  For
example, scale invariant density fluctuations with mass typical of
dwarf galaxies ($10^8-10^9$ M$_\odot$) grow to an amplitude $\delta
\sim 10^{-2}$ by redshift of matter-radiation equality. They collapse
when $\delta \sim 1.68$ at $z_{vir} \sim 15$ following further growth
by a factor $z_{eq}/z_{vir} \sim 200$ during matter-domination era.

Dark matter perturbations that enter the Horizon with amplitude
$\simgt 0.1\%$ -- \ie, $\sim 100$ times larger than the approximately
scale invariant perturbations from inflation -- collapse before
$z=1000$. These perturbations are too small to form PBHs but can seed
the growth of ultracompact minihalos that accrete from a nearly
uniform Universe, before the epoch of formation of the first galaxies
at $z \sim 30$. Thus, the angular momentum of the accreted dark matter
is small and the infall quasi-radial.

The aforementioned arguments lead us to conclude that the formation of
ultracompact minihalos is more likely than the formation of PBHs. The
threshold overdensity required for their formation at $z \simgt 1000$
is $\sim 0.1\%$, compared to $\approx 30\%$ required for PBH
formation.

Similarly to the case for PBHs, observational constraints or the
discovery of ultracompact minihalos is a powerful probe of high-energy
physics in the early Universe. Microlensing experiments can help
constrain the amplitude of perturbations produced during two recent
phase transitions: the QCD (quark-hadron) phase transition with
$M_{Hor}\sim 1$~M$_\odot$ and the $e^+-e^-$ annihilation epoch with
$M_{Hor} \sim 10^5$~M$_\odot$. Perturbations with masses in this range
can be constrained by microlensing experiments because the mass of the
lens, given by equation~(\ref{eq:halolens}), is in the range probed by
current microlensing experiments.

\subsection{Angular Momentum of Accreted Dark Matter}

The angular momentum of accreted dark matter determines the density
profile of the dark halo enveloping a PBH and whether the mass of a
PBH can grow substantially by accreting a fraction of its
enveloping dark halo. ROM09 found that angular momentum of dark matter
accumulated by PBHs with masses $M_{pbh}>1000$ is so small that a
fraction of the dark matter is directly accreted by the PBH. But for
smaller mass PBHs and ultracompact minihalos the angular momentum is
larger and the accretion can become non radial in the inner parts of
the halo.  The $\nu=2.25$ power law profile of the density of
minihalos results from the assumption of radial infall. If the
accreted material has some angular momentum the density profile will
be shallower near the center and will contain less mass. Here we check
whether the assumption of radial infall is justified in the inner
parts of the halo profile enclosing a mass comparable to the lensing
mass (\ie, within the Einstein radius).
 
The angular momentum of the dark matter accreting onto PBHs can be
estimated from equation~(19) in ROM09, which gives the mean values of
the velocity dispersion within a comoving volume of radius equal to
the halo turnaround radius,
\begin{equation}
\sigma_{dm} \approx \sigma_{dm,0} \left({1+z \over 1000}
\right)^{-{1\over2}}\left({M_h \over1~M_\odot}\right)^{0.28},
\end{equation}
with $\sigma_{dm,0}=1.4 \times10^{-4}~{\rm km~s}^{-1}$.

Applying conservation of angular momentum we find that the rotational
(i.e., tangential) velocity of the gas at a distance $r$ from the
black hole is $v(r)r=\sigma_{dm}R_h$, where $R_h$ is the halo radius.
If the velocity is smaller than the Keplerian velocity in the
proximity of the Einstein radius, then the accretion is
quasi-spherical; if vice versa, a shallower core can form. The
Keplerian velocity at radius $R$ is $(GM(R)/R)^{1/2}$. Thus, the
accretion is quasi-spherical within the Einstein radius if
$\sigma_{dm}R_h< (GM_{lens}R_E)^{1/2}$. This inequality is satisfied
for
\begin{equation}
\delta m < 1350 M_\odot \left({1+z \over 1000}
\right)^{3.46}\left({M_{lens} \over 1~M_\odot}\right)^{1.22}.
\label{eq:ang1}
\end{equation}
In equation~(\ref{eq:ang1}) there is a steep dependence on redshift. A
value of redshift $z \sim 1000$ is most appropriate for probing the
inner parts of the density profile because the inner mass is dominated
by material accreted at high redshift.  \citet{Bertschinger:85} has
shown that the growth of the halo is self similar because the material
accreted at late times from outer shells does not change the density
profile in the inner parts (despite the fact that these carry most of
the halo mass).  This is because the outer shells have high speed when
they reach the halo center and spend little time there. Thus, the mass
profile near the center of the minihalo - within the Einstein radius -
is built by material accreted at redshifts $z \simgt 1000$.  Let us
now rewrite equation~(\ref{eq:ang1}) for two special cases:
\begin{itemize}
\item[1)] If $M_{lens} \approx M_{pbh}$, assuming $\delta m=M_{pbh}/f_{Hor}$ we find that the infall is quasi-radial at $R_E$ if
\begin{equation}
7.4 \times 10^{-4} \left({1+z \over 1000}
\right)^{-3.46}\left({M_{pbh} \over 1~M_\odot}\right)^{-0.22} <f_{Hor}<1.
\end{equation}
\item[2)] If $M_{lens} \approx M_{\perp}$ from equation~(\ref{eq:halolens}) we
  find that the infall is quasi-radial at $R_E$ if
\begin{equation}
\delta m > 2.9 \times 10^{-2}~M_\odot \left({1+z \over 1000}\right)^{-7.4}.
\end{equation}
\end{itemize}
% 1pc=2e5* AU

\subsection{Profile Steepening due to Adiabatic Compression and \pop3
  Formation}

In this section we discuss the role of baryons on shaping the density
profile of ultracompact minihalos and whether the first \pop3 stars
can form in their centers. The discussion here will be only
qualitative as the simulations required to obtain quantitative results
are beyond the scope of the present paper.

\begin{figure}[thp]
\plotone{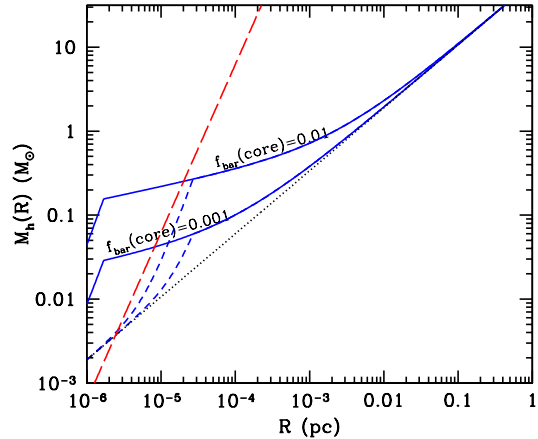}
\caption{\label{fig:lens} Dark matter mass profile of minihalos before
  (dotted line) and after (solid and dashed lines) baryonic
  compression. The initial mass profile is for a minihalo at $z=10$
  with $\delta m=1$~M$_\odot$. We show the mass profile for a set of
  final baryon configurations. For simplicity here we assume that the
  baryons have a constant density core with radius $R_{core}=0.5$~AU
  (solid lines) and $8$~AU (dashed lines) and with gas mass that is a
  small fraction $f_{bar}(core)=0.01, 0.001$ of $M_h$ (see labels).}
\end{figure}

Due to gas cooling and the concentrated dark matter profile of
ultracompact minihalos, baryons may sink dissipatively by several
orders of magnitude in radius toward the halo center. The infall
perturbs the underlying dark matter distribution pulling it inward and
creating an even smaller and denser core than would have evolved without
baryon condensation. For the case of ultracompact minihalos, it is
most appropriate to assume perfectly radial orbits (but the same
equations are valid assuming only circular orbits). Since $M_h(R)$
varies in a self-similar fashion, the quantity $R_{max}M_h(R_{max})$,
where $R_{max}$ is the maximum radius of the radial orbit, is
conserved during the ``slow'' contraction of the gas
component. Defining the fraction of baryonic mass $F \equiv M_b/M_h$,
the invariant of the dark matter particle orbits implies
\begin{equation}
R[M_b(R)+M_{h}(R)]=R_i M_i(r_i)={R_iM_{h,i}(R) \over 1-F}
\label{eq:compr}
\end{equation}
which can be solved iteratively for the final dark matter mass
distribution $M_{h}(R)$, given the initial total mass distribution
$M_{h,i}(R_i)$ and the final baryon mass distribution $M_b(R)$
\citep{Blumenthal:86}. The equality $M_{h,i}=(1-F)M_i(R_i)$, further
assumes that initially $F(R_i)=const$.  In Figure~\ref{fig:lens} we
show the compression of the dark matter profile due to baryonic
contraction obtained by solving iteratively \eq~(\ref{eq:compr}). The
initial minihalo mass profile is a power law $M_{h,i}=M_h(z)
(R/R_h)^{0.75}$ with $M_h(z)$ from equation~(\ref{eq:mhalo_new})
assuming $\delta m =1$~M$_\odot$ and $z=10$ (dotted line). The final
dark matter profiles are shown by the solid and dashed lines for
different configurations of the final baryon distribution. We assume
that a fraction of the baryons $f_{bar}(core)=0.01, 0.001$ settles to
a constant density core with radii $R_{core}=0.5$ AU and $8$ AU.  The
value of $R_{core}$ is not important as long as $R_{core}<R_E$. The
mass of the lens is given by the intersection of the final dark matter
mass profile with the long dashed line, which traces the lens equation
$M_{lens}/(1\,M_\odot) \approx [R_E/(8 AU)]^2$. The mass of lens
increases only if the gas mass within $R_E$ for the initial mass
profile is comparable or dominates initial dark matter mass. If later,
the baryons are removed from the center of the halo on a short time
scale when compared to the orbital period of the dark matter particles
(for instance due to reionization, SN explosions, etc) the cuspy dark
matter profile may become less steep, but it will not go back to the
original profile before the baryon compression \citep{GnedinK:04,
  Sellwood:05}.

The next step consists in determining whether a fraction of the
baryons in the minihalo can become sufficiently concentrated to
dominate the mass within $R_E$ at some point during the minihalo
evolution. We will consider separately the case of PBHs embedded in
minihalos and ultracompact minihalos without PBH.

\noindent
1) For PBHs accreting at high-z from the IGM, the gas mass within the
Einstein radius of the lens is proportional to the dimensionless gas
accretion rate ${\dot m}={\dot M_g}/{\dot M_{Ed}}$, where $\dot M_{Ed}$
is the Eddington rate. For PBHs with mass $\simlt 500$~M$_\odot$ the
gas accretion is quasi-radial. Thus, assuming $M_{lens} \sim M_{pbh}$,
the gas approaches free fall velocity $v \sim
(GM_{lens}/R)^{1/2}$. From mass conservation $4\pi \rho(R) R^2v={\dot
  m}{\dot M_{Ed}}$, we find that
\begin{equation} 
{M_g(R_E) \over M_{lens}}\sim 5.4\times 10^{-8}{\dot
  m}\left({M_{lens} \over 1~M_\odot}\right)^{1/4}.
\label{eq:mgas}
\end{equation}
Thus, in this regime, the gas mass never dominates the potential
within $R_E$.  However, equation~(\ref{eq:mgas}) does not apply in the
following cases: i) PBHs more massive than $M_{pbh}>10^5$~M$_\odot$,
for which the gas is self-gravitating \citep{Ricotti:07}; ii) PBHs
with mass $>500-1000$~M$_\odot$ accreting gas at $z<100$, for which
the gas has sufficient angular momentum to form an accretion
disk. Thermal feedback effects, which are important at $z < 100$, will
tend to increase the value of the PBH critical mass for disk formation
(ROM08).  An accretion disk may also form around PBHs that are able to
accrete gas efficiently from the ISM in galaxies. However, to sustain
efficient gas accretion the relative velocity of PBHs with respect to
the ISM must be subsonic. Only in this latter case, if the accretion
rate is near Eddington, the accretion disk may dominate the mass
profile within $R_E$.  We conclude that compression of dark minihalos
around PBHs due to baryon dissipative collapse is in most cases
negligible.

\noindent
2) The first stars are thought to form in rare $3\,\sigma$
perturbations, of mass $\sim 10^5$~M$_\odot$ at $z\sim 30$
\citep{Tegmark:97}. Ultracompact minihalos, because they are already
in place at $z\sim 1000$, may form the first stars in smaller mass
halos of $10^4$~M$_\odot$ at redshifts $200 \simlt z \simlt 1000$. The
simplest and naive condition for star formation is given by the
requirement that the minihalos must have $T_{vir}>T_{cmb}$: the gas
would not be able to condense if it were initially colder than the
temperature of the CMB (see Fig~6 in \citealt{Tegmark:97}).
%The circular velocity of ultracompact minihalos is about a factor of 2
%smaller than $v_{cir}$ at virialization of normal minihalos.
Thus, ultracompact minihalos with $\delta m > 100$~M$_\odot$ are able
to host the formation of \pop3 stars, and the baryon dissipation can
increase the concentration of the dark matter halo.

%Depending on the minihalo mass, the gas can cool to $T<T_{igm}$ due to
%H$_2$ cooling at $z\simlt 200$.
However, the naive estimate is likely too conservative because
ultracompact minihalos have a power law density profile with
$\nu=2.25$, thus are much more dense than normal galaxies (\eg,
assuming NFW profile). Their steep mass profile leads to significantly
larger gas densities in their cores. Assuming an isothermal equation
of state with $T_g=T_{CMB}$ (a good approximation at $z>100$ due to
Compton cooling/heating), the gas density profile in hydrostatic
equilibrium in the potential of an ultracompact minihalo is
\begin{equation}
\rho_b(r)=\overline \rho_b \exp\left\{\alpha\left[\left({R \over R_h}\right)^{-1/4}-1\right]\right\}
\label{eq:beta}
\end{equation}
where $\alpha=4 r_b/r_h$ \citep{Ricotti:09}. For $T_{g}=T_{CMB}$ we get
$\alpha=(M_h/240~M_\odot)^{2/3} \approx 0.054 \delta m^{2/3}
(1+z/1000)^{-{2/3}}$. If we estimate $\rho_b(R)$ at $R=R_E$, where
\begin{equation}
{R_E \over R_h}=1.6 \times 10^{-5} \left({\delta m \over 1~M_\odot}\right)^{13/15}\left({1+z \over 1000}\right)^{4/3},
\end{equation}
we find that the core gas density is
\begin{equation}
  \rho_b(R_E) \sim \overline \rho_b \exp{\left\{\alpha\left({R \over R_h}\right)^{-1/4}\right\}} \sim \overline \rho_b 10^{0.37 ({\delta m \over 1~M_\odot})^{0.45}({1+z \over 1000})^{-1}}.
\label{eq:gcore}
\end{equation}
The gas density in the core becomes increasingly large with decreasing
redshift.  Equation~(\ref{eq:gcore}) is a good estimate of the gas
density in the core as long as the gas temperature is close to or
lower than the CMB temperature (\ie, $z>10-30$) and
$\rho_b(R_E)<\rho_{dm}(R_E)$.

If the gas in the core is able to form H$_2$ and cools in less than a
Hubble time, it may collapse and form the first \pop3 stars. The large
initial density in the core increases the rate of formation of H$_2$
with respect to the estimates by \citet{Tegmark:97} for normal
galaxies (although the reduced fractional ionization and $H^-$
abundance in the high density core have the opposite effect, these do
not dominate). Thus, ultracompact minihalos of mass significantly
smaller than $10^4$~M$_\odot$ at $z \sim 30$ may be able to form dense
gas cores or \pop3 stars. It is therefore possible that minihalos with
$\delta m \simlt 1$~M$_\odot$, as the one shown in
\fig~\ref{fig:lens}, may be compressed by baryon collapse and produce
lenses with much larger masses than given by
equation~(\ref{eq:halolens}). Quantitative calculations of the gas
collapse require the integration of a set of time-dependent equations
of the chemical reaction network for H$_2$, in the evolving
gravitational potential of the minihalo. These calculations are beyond
the scope of the present paper.
%At any redshift the mean overdensity of minihalos within the radius
%$R$ is $\delta(R)=10 (R/R_h)^{0.75-3}$. 

\section{Microlensing Light Curve}\label{sec:mag}

\begin{figure*}[tbh]
\epsscale{1.0}
\plottwo{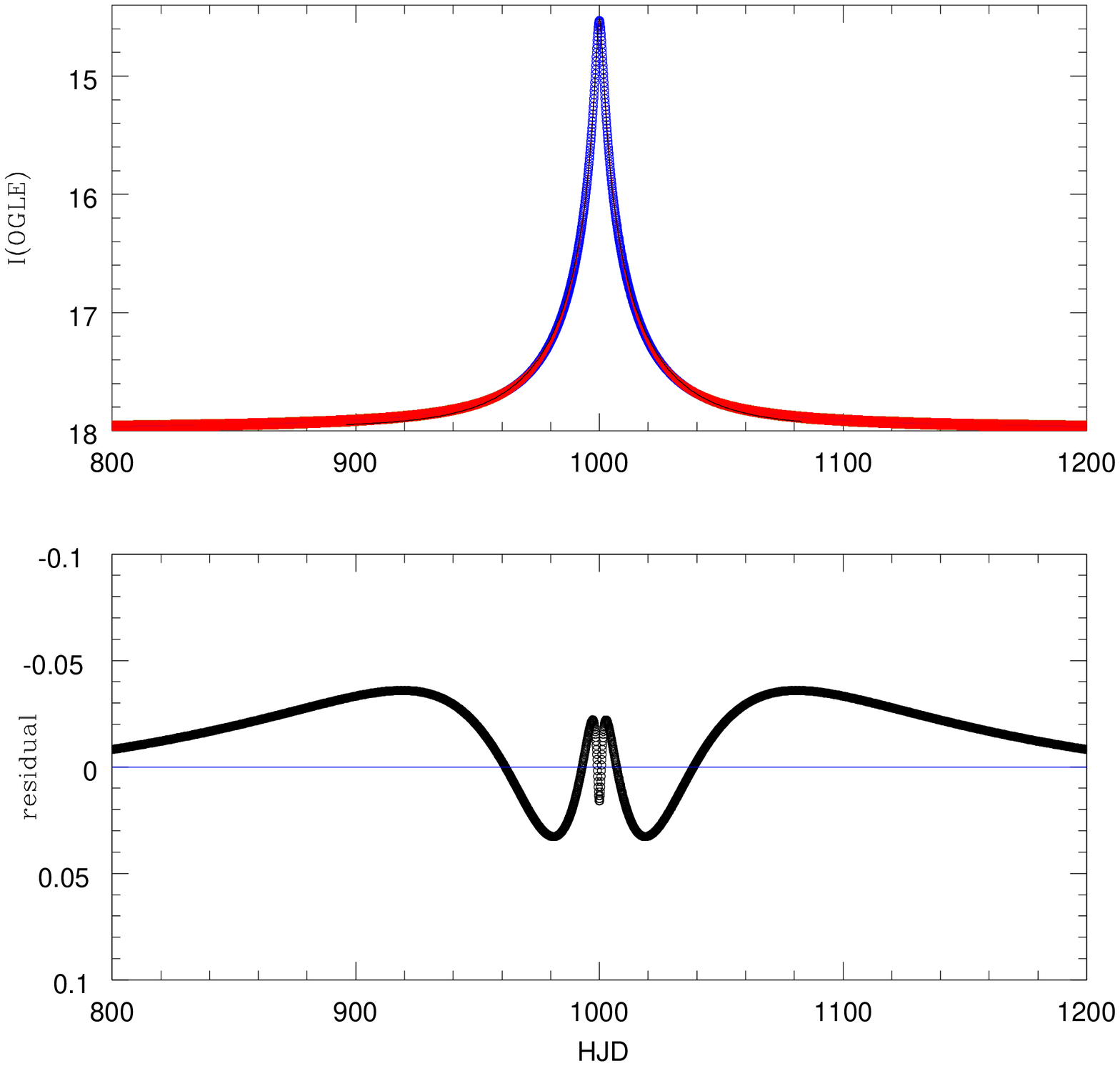}{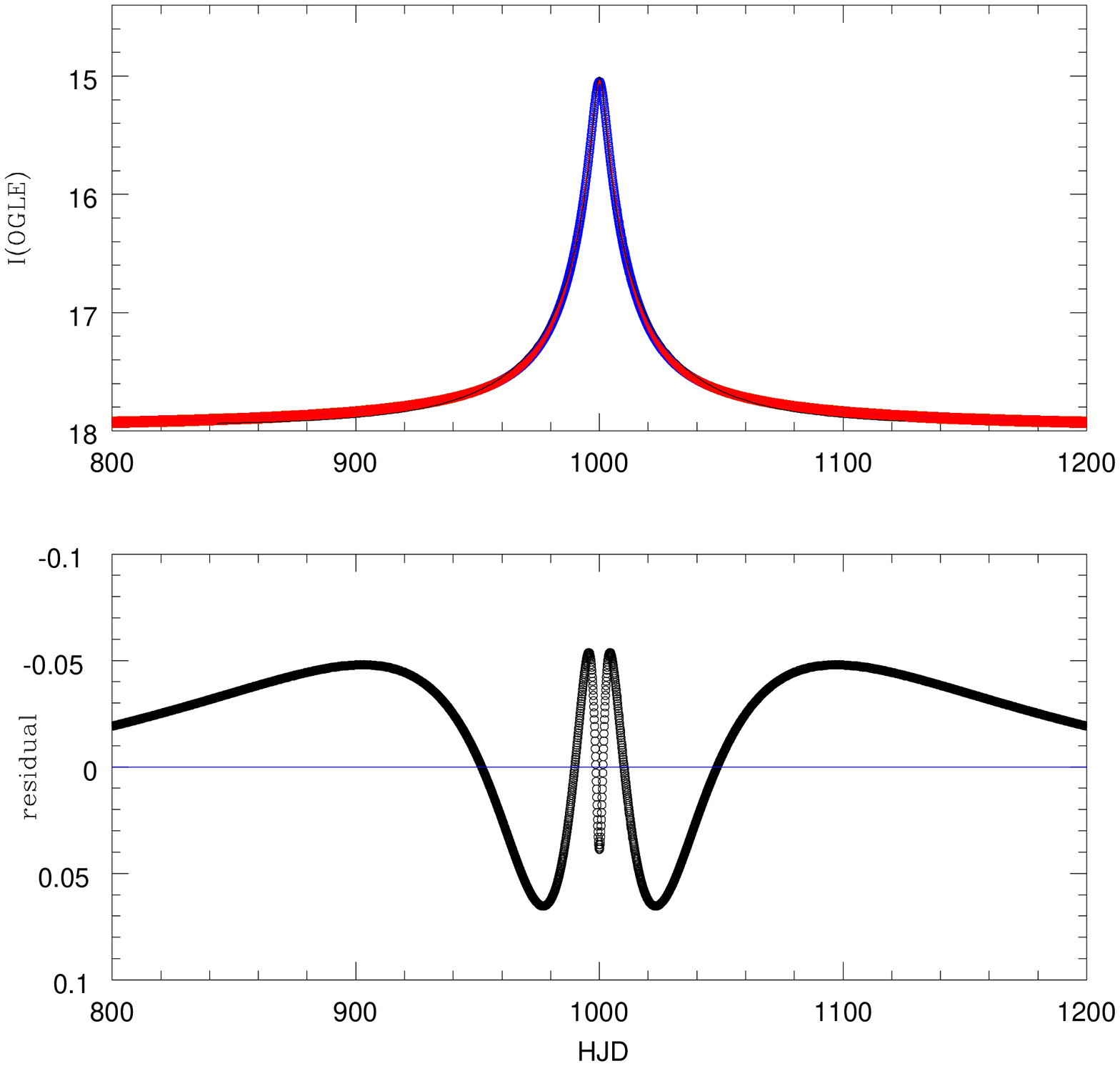}
\caption{\label{fig:mag} {\it (Left).} Magnification light curve (top
  panel) of a PBH of $0.5$~M$_\odot$ including its minihalo with mass
  within $r_E=8 AU$ of $1$~M$_\odot$. Here the source impact
  parameter and crossing time in units of the Einstein radius are
  $0.05$ and $30$ days, respectively. The bottom panel shows the
  residual with respect to the magnification curve of a
  point mass. The synthetic data were taken for 10 Einstein ring
  crossing times (about 10 months in this example).{\it (Right).} Same
  as the left figure for an ultracompact minihalo (without PBH) with
  $M_{lens}=1$~M$_\odot$.}
\label{fig:mag}
\end{figure*}

To calculate the magnification of minihalo MACHOs (initially without
PBHs at their center), we consider density profiles of the form $\rho
= C r^{-v}$, where $C$ is set so that the {\it projected mass}
$M_\perp$ within a radius $R_{\rm E}$ is just equal to mass required
to form an Einstein ring at $R_{\rm E}$.  That is,
\begin{eqnarray}
M_\perp(R) &=& \int_0^R 2\pi db\,b\int_{-\infty}^{\infty}
dz\, C\,(b^2 + z^2)^{-\nu},\\
R_{\rm E}^2 &=& {4 G M_\perp(R_{\rm E})\over c^2}D_{\rm L}^2
\biggl({1\over D_{\rm L}} - {1\over D_{\rm S}}\biggr)
\end{eqnarray}
where $D_{\rm L}$ is the distance from Earth to the minihalo lens and
$D_{\rm S}$ is the distance to the microlensed source.  Note that
$M_\perp(R)$ is related to the mass inside a sphere of radius $R$ by
\begin{equation}
{M_\perp(R)\over M(R)} =
{(1/2)![(\nu-3)/2]!\over[(\nu-2)/2]!}
\label{eq:proj1}
\end{equation}
which is 1.350 for $\nu=2.25$.  Hence, $M_\perp(R)\propto R^{3-\nu}$.

We then use Newton's method to solve the lens equation, which gives
the image position $\theta_I$ as an implicit function of the source
position $\theta_S$,
\begin{equation}
\theta_I - \theta_S = {4 G M_\perp(D_L\theta_I)\over \theta_I c^2}
\biggl({1\over D_{\rm L}} - {1\over D_{\rm S}}\biggr).
\end{equation}
Newton's method automatically returns the radial derivative of
$M_\perp$, and so allows calculation of the magnification,
\begin{equation}
A = A_+ + A_-; \qquad A_\pm =
\bigg|{\partial\theta_I\over\partial\theta_S}{\theta_I\over\theta_S}\bigg|.
\end{equation}

For the case that the minihalo is seeded by a PBH, we simply add a
point mass to the center of the halo profile and repeat the above
procedure.

We construct simulated light curves (with errors proportional to the
square root of the magnification) and proceed to fit the simulated
flux $F$ to a standard 5-parameter point-lens microlensing model of
the form
\begin{eqnarray}
F &=& f_s A(t;t_0,u_0,t_{\rm E}) + f_b\\
A(u) &=& {u^2 + 2\over u\sqrt{u^2+4}};
\qquad u(t) = \sqrt{u_0^2 + {(t-t_0)^2\over t_{\rm E}^2}}
\label{eq:plens}
\end{eqnarray}
where $t_{\rm E}$ is the Einstein radius crossing time, $u_0$ is the
impact parameter in units of the Einstein radius, $t_0$ is the time of
maximum (closest approach), $f_s$ is the model source flux, and $f_b$
is background light that lies in the photometric aperture but does not
participate in the event.  We adopt the conservative approach of
allowing negative blending (which would be unphysical) for two
reasons. First, the great majority of microlensing events are blended
because they are in crowded fields.  This blended light typically
absorbs any negative blending from a spurious fit, rendering the
latter undetectable.  Second, even truly isolated sources can suffer
negative blending because of poor determination of the sky.

The two examples in Figure 3 show deviations from standard
microlensing of somewhat different degree.  Whether either of these
would be detectable in practice would depend on the quality of the
experiment.  However, the crucial point is that {\it they are
  detectable just based on their structure}.  When microlensing dark
matter searches were proposed by Paczy\'nksi (1986), their major
selling point was that if the dark matter were composed of MACHOs,
then microlensing events would be 25 times more common than if it were
not.  Hence the event rate itself would unambiguously settle the
question.  Then, when MACHO reported an event rate that was the
root-mean-square of the prediction for MACHOs and the prediction for
stars, the result became completely ambiguous.  Recent work by EROS
and OGLE has seemed to show that the event rate is compatible with
that expected from stars.  If confirmed, this would have seemed to be
the end-game for microlensing dark-matter searches: even if there were
dark-matter objects, they would be too few to detect as an excess
event rate (without enormously larger experiments) and, in any event,
would be cosmologically uninteresting because their density was so far
below the dark-matter density.

But Figure 3 shows that minihalo MACHOs have light curve signatures
that are different from those of stars and therefore can be detected
even if they are only as common as stars, or indeed, even if they are
far less common.  Moreover, as we have discussed in
Section~\ref{sec:int}, minihalo MACHOs could make up all the dark
matter and yet have a microlensing optical depth that is one or
several orders of magnitude below that expected for point-lens dark
matter.

We note that there are several observed microlensing effects that
produce deviations from the standard microlensing form given by
equation~(\ref{eq:plens}).  For example, parallax effects due to the
Earth's acceleration, xallarap effects due to source acceleration from
an orbiting companion, light curve caustics due to binary or planetary
companions, and finally finite-source effects that occur if the lens
transits the source.  However, except for the last, all of these are
generically asymmetric, whereas the light curves shown in Figure 3 are
symmetric.  Moreover, although finite-source effects are symmetric,
they have a very specific form which is unlike the form due to
minihalo MACHOs.  Furthermore, they are almost completely restricted
to extremely high-magnification events, which are rare.

Now it is true that the other effects could, in particular cases, be
very symmetric but this is highly improbable, and furthermore these
effects, even in their more typical asymmetric form, are not all that
common.  Hence, the level of contamination in minihalo-MACHO searches
would be extremely low.

\section{MACHO Searches Toward the Bulge}\label{sec:bulge}

Originally, \citet{Pac:86} proposed MACHO searches toward the LMC not
because this line of sight had the highest expected number of MACHO
microlensing events, but because it had very little contamination from
non-MACHO events.  However, if the MACHO and non-MACHO events can be
distinguished based on their light curves, one should really focus
MACHO searches on the line with the highest rate of dark-halo events
and the highest-quality light curves (needed to decisively
characterize the deviations from point-lens microlensing).  This is
the Galactic bulge, where currently about 800 events are discovered
per year and where the sources are 6 times closer, so 36 times
brighter (for similar luminosities).  It is true that the fraction of
these due to MACHOs is only $0.6\% (f/0.1)$, where $f$ is the ratio of
MACHO optical depth toward the LMC to that expected for a full
point-lens MACHO halo \citep{Gould:05}.  However, these $\sim 5
(f/0.1)$ per year events are to compared to the $\sim 1$ per year
currently detected toward the MCs.

Of course, the problem of background rejection is much more severe
because there are $\sim 10^3$ times more stellar lensing events.  But
the quality of the lightcurves is also higher, not only because the
sources are closer but because the bulge fields are much more
aggressively monitored than the MCs, due to the possibility of finding
extra-solar planets.  This will be even more true in the future, as
next generation microlensing planet-search experiments (already
funded) begin to come on line.

And further in the future, there are proposals to put a wide-field
imager in space, which could obtain the exquisite photometry that
would enable one to distinguish even extremely subtle differences
between point-lens and compact-minihalo microlensing events.  Such a
satellite would already create a synergy between dark-energy and
extra-solar planet studies \citep{Gould:09}, and its additional power
to probe for compact minihalos (and so the early-universe phase
transitions that generate them) would add both dark-matter searches
and early-universe physics to the mix.

\section{Conclusions and Discussion}\label{sec:conc}

We propose a new type of dark matter MACHO that may form copiously in
the early Universe. Upper limits on, or the discovery of these MACHOs
is a powerful new tool to study cosmic phase transitions and the
nature of dark matter. The new MACHOs are ultracompact minihalos
that may form from cosmological perturbations in the Universe that are
only 10-100 times larger than scale-invariant perturbations from
inflation that seeded the formation of large scale structures and
galaxies.  These objects are more likely to form than PBHs, due to the
much larger amplitude of perturbations that are required to form
PBHs. Ultracompact minihalos may be sites of formation of the first
\pop3 stars. We show that the magnification light curve of
ultracompact minihalos is similar to the one of a point mass but can
be easily recognized from it by sampling the light curve for a
sufficiently long time after the peak of the magnification event.

Similarly, if PBHs exist and are only a fraction of the dark matter,
their light curve would differ from those due to point masses because
PBHs seed the growth of an enveloping minihalo that is sufficiently
massive and concentrated to modify the microlensing magnification
curve.  This modification is substantial whenever the mass of the
minihalo within the lens Einstein radius is $\sim 30\%$ of the PBH
mass. However, the PBH mass can be much smaller than the minihalo and
MACHO masses.

For the mass profile of the minihalo-enveloping PBHs as predicted by
MOR07, the mass within the Einstein radius is only $3\%$ of the PBH
mass, thus the PBH magnification curve does not show measurable
modifications. However, MOR07 assumed secondary infall from an
expanding uniform Universe seeded by a point-mass PBH. The later
assumption in most cases is incorrect: PBHs form from collapse of
linear fluctuations on Horizon scales, but their mass is typically
smaller than the Horizon mass: $M_{PBH}=f_{Hor}M_{Hor}$, with
$f_{Hor}<1$. Thus, the mass that seeds the secondary infall is larger
than the PBH mass. This leads to more massive and more compact
minihalos enveloping PBHs, which have observable signatures on the
microlensing light curves.

The mass of the lens produced by compact minihalos scales with the
mass of the perturbation as $\delta m^{6/5}$. Thus, statistical
fluctuations of the mass of the perturbations $\delta m$ produce a
mass range of the gravitational lenses (\ie, a range of MACHO
masses). This is also the case for minihalos surrounding PBHs, if they
exist. In this second case we expect variations of the ratio of MACHO
masses to PBH masses.

Prior to this work, microlensing searches for PBHs were reaching a
dead-end.  Although the MACHO experiment \citep{Alcock:00} had
initially found a microlensing signature toward the MCs that was too
large to be produced by known populations of stars, and indeed was
difficult to explain by any objects other than PBHs, subsequent
results reported by EROS \citep{Tisserand:07} and OGLE
\citep{Wyrzykowski:09} have tended to cast doubt on MACHO's claim.
But even if these are ultimately verified, ROM08 have argued that the
density of PBHs required to explain the MACHO results would give rise
to other signatures that have not been observed.  This seemingly led
to limits of order 1\% on PBHs, far below the reported MACHO value,
but also (and more importantly), well below the background level
expected from Galactic and MC stellar microlenses.  Hence, it had
appeared as though PBHs were inaccessible to microlensing experiments
and, in any event, cosmologically unimportant (even if perhaps
interesting).

However, we have shown that PBH-seeded minihalos remain viable
candidates for the origin of the signal reported by MACHO, and
further, that even if this reported signal proves spurious, it is
still possible to search for PBHs in microlensing data at lower levels
(well below the previously conceived ``floor'' of the stellar
background).  Both these realizations derive from the fact that PBHs
make up only a small part of the mass of PBH-seeded minihalos.  Hence,
the cosmological density of PBHs can be low while the halos they seed
can generate substantial microlensing signal.  And the fact that
minihalo+PBHs are not point-like leads to a qualitatively different
microlensing signal.

Moreover, we have shown that the same type of early-universe
perturbations that give rise to PBHs, will (at lower amplitude) give
rise to minihalos that lack PBHs.  Such minihalos are both more likely
(because they are produced by a larger range of initial conditions)
and easier to detect (because they deviate more strongly from
point-lens profiles).  Therefore microlensing experiments (past and
future) are a far more powerful probe of early-universe physics than
previously understood.

Probing the clumpiness of dark matter at mass scales accessible to
microlensing experiments has several implications to understand the
nature of dark matter and the high-energy Universe:
\begin{enumerate}
\item It provides the best
  test to determine whether the dark matter is cold or warm. The
  masses of thermal or non-thermal relics and their free-streaming
  length could be constrained much better than in studies based on the
  power spectrum at small scales from the Lyman-$\alpha$ forest and
  the number of satellites in the Milky Way.
\item It provides information on the physics of the high-energy
  Universe by constraining the amplitude of inhomogeneities created at
  phase transitions due to bubble nucleation, topological defects
  formation and other non-Gaussian processes.
\item The clumpiness of dark matter in the Milky Way halo affects the
  flux of dark matter particles on earth, important for detections by
  ground based experiments. Unless a clump happens to intersect the
  earth, the average dark matter density on earth would be reduced
  with respect to the standard value $\sim 0.008$~M$_\odot$pc$^{-3}$.
\item The centers of ultracompact minihalos may be powerful emitters
  of gamma rays due to self-annihilation of WIMPS. A test of our model
  would consist in follow up observations of gamma rays toward the
  line of site of the magnification event.
\item Ultracompact minihalos may have been the sites of formation of
  the first stars formed in our Universe.
\end{enumerate}
A positive identification of PBHs as MACHOs would motivate future
space missions to detect signatures of the energy injection by PBHs in
the early Universe (ROM08). The best measurements of the CMB spectrum
to date are by FIRAS on board of COBE, now 15 years old. A follow up
mission would be of great scientific value and easily justified if
evidence is found of the existence of non-baryonic MACHOs.  Early
energy injection also modifies the cosmic recombination history and
thus the spectrum of anisotropies of the CMB. The Planck mission will
improve existing upper limits on PBHs and may detect signatures of
their existence.

\acknowledgements 
Work by MR was supported by the Theoretical Astrophysics program at
the University of Maryland (NASA grant NNX07AH10G and NSF grant
AST-0708309).  Work by AG was supported by NSF grant AST-0757888.

\bibliographystyle{/Users/ricotti/Latex/TeX/apj}
\bibliography{/Users/ricotti/Latex/TeX/archive}

\begin{thebibliography}{}

\bibitem[\protect\citeauthoryear{{Ahmed} et~al.}{{Ahmed}
  et~al.}{2009}]{WIMP:09}
{Ahmed}, Z., et~al. 2009, Physical Review Letters, 102, 011301

\bibitem[\protect\citeauthoryear{{Alcock} et~al.}{{Alcock}
  et~al.}{2000}]{Alcock:00}
{Alcock}, C., et~al. 2000, \apj, 542, 281

\bibitem[\protect\citeauthoryear{{Angle} et~al.}{{Angle}
  et~al.}{2008}]{WIMP:08}
{Angle}, J., et~al. 2008, Physical Review Letters, 100, 021303

\bibitem[\protect\citeauthoryear{{Bertschinger}}{{Bertschinger}}{1985}]{Bertsc%
hinger:85}
{Bertschinger}, E. 1985, \apjs, 58, 39

\bibitem[\protect\citeauthoryear{{Blumenthal} et~al.}{{Blumenthal}
  et~al.}{1986}]{Blumenthal:86}
{Blumenthal}, G.~R., {Faber}, S.~M., {Flores}, R.,  \& {Primack}, J.~R. 1986,
  \apj, 301, 27

\bibitem[\protect\citeauthoryear{{Carr}}{{Carr}}{1975}]{Carr:75}
{Carr}, B.~J. 1975, \apj, 201, 1

\bibitem[\protect\citeauthoryear{{Carr}}{{Carr}}{2005}]{Carr:05}
{Carr}, B.~J. 2005, ArXiv Astrophysics e-prints

\bibitem[\protect\citeauthoryear{{Carr} \& {Hawking}}{{Carr} \&
  {Hawking}}{1974}]{CarrH:74}
{Carr}, B.~J.,  \& {Hawking}, S.~W. 1974, \mnras, 168, 399

\bibitem[\protect\citeauthoryear{{Chisholm}}{{Chisholm}}{2006}]{Chisholm:06}
{Chisholm}, J.~R. 2006, \prd, 73, 083504

\bibitem[\protect\citeauthoryear{{Dokuchaev}, {Eroshenko}, \&
  {Rubin}}{{Dokuchaev} et~al.}{2004}]{Dokuchaev:04}
{Dokuchaev}, V., {Eroshenko}, Y.,  \& {Rubin}, S. 2004, ArXiv Astrophysics
  e-prints

\bibitem[\protect\citeauthoryear{{Gnedin} et~al.}{{Gnedin}
  et~al.}{2004}]{GnedinK:04}
{Gnedin}, O.~Y., {Kravtsov}, A.~V., {Klypin}, A.~A.,  \& {Nagai}, D. 2004,
  \apj, 616, 16

\bibitem[\protect\citeauthoryear{{Gould}}{{Gould}}{2005}]{Gould:05}
{Gould}, A. 2005, \apj, 630, 887

\bibitem[\protect\citeauthoryear{{Gould}}{{Gould}}{2009}]{Gould:09}
{Gould}, A. 2009, Astro2010: The Astronomy and Astrophysics Decadal Survey,
  Science White Papers, no. 100, ``Wide Field Imager in Space for Dark Energy
  and Planets'', (arXiv:0902.2211)

\bibitem[\protect\citeauthoryear{{Green} et~al.}{{Green}
  et~al.}{2004}]{Green:04}
{Green}, A.~M., {Liddle}, A.~R., {Malik}, K.~A.,  \& {Sasaki}, M. 2004, \prd,
  70, 041502

\bibitem[\protect\citeauthoryear{{Mack}, {Ostriker}, \& {Ricotti}}{{Mack}
  et~al.}{2007}]{MackOR:07}
{Mack}, K.~J., {Ostriker}, J.~P.,  \& {Ricotti}, M. 2007, \apj, 665, 1277

\bibitem[\protect\citeauthoryear{{Paczynski}}{{Paczynski}}{1986}]{Pac:86}
{Paczynski}, B. 1986, \apj, 304, 1

\bibitem[\protect\citeauthoryear{{Ricotti}}{{Ricotti}}{2007}]{Ricotti:07}
{Ricotti}, M. 2007, \apj, 662, 53

\bibitem[\protect\citeauthoryear{{Ricotti}}{{Ricotti}}{2009}]{Ricotti:09}
{Ricotti}, M. 2009, \mnras, 392, L45

\bibitem[\protect\citeauthoryear{{Ricotti}, {Ostriker}, \& {Mack}}{{Ricotti}
  et~al.}{2008}]{RicottiOM:08}
{Ricotti}, M., {Ostriker}, J.~P.,  \& {Mack}, K.~J. 2008, \apj, 680, 829

\bibitem[\protect\citeauthoryear{{Sellwood} \& {McGaugh}}{{Sellwood} \&
  {McGaugh}}{2005}]{Sellwood:05}
{Sellwood}, J.~A.,  \& {McGaugh}, S.~S. 2005, \apj, 634, 70

\bibitem[\protect\citeauthoryear{{Tegmark} et~al.}{{Tegmark}
  et~al.}{1997}]{Tegmark:97}
{Tegmark}, M., {Silk}, J., {Rees}, M.~J., {Blanchard}, A., {Abel}, T.,  \&
  {Palla}, F. 1997, \apj, 474, 1

\bibitem[\protect\citeauthoryear{{Tisserand} et~al.}{{Tisserand}
  et~al.}{2007}]{Tisserand:07}
{Tisserand}, P., et~al. 2007, \aap, 469, 387

\bibitem[\protect\citeauthoryear{{Wyrzykowski} et~al.}{{Wyrzykowski}
  et~al.}{2009}]{Wyrzykowski:09}
{Wyrzykowski}, {\L}., et~al. 2009, \mnras, 890

\bibitem[\protect\citeauthoryear{{Zel'Dovich} \& {Novikov}}{{Zel'Dovich} \&
  {Novikov}}{1967}]{Zeldovich:67}
{Zel'Dovich}, Y.~B.,  \& {Novikov}, I.~D. 1967, Soviet Astronomy, 10, 602

\end{thebibliography}

\end{document}